\documentstyle[12pt]{article}
\newcommand{\be}{\begin{equation}}
\newcommand{\ee}{\end{equation}}
\newcommand{\bea}{\begin{eqnarray}}
\newcommand{\eea}{\end{eqnarray}}
\newcommand{\brr}{\begin{array}}
\newcommand{\prd}{Phys.Rev. \underline}
\newcommand{\pl}{Phys.Lett. \underline}
\newcommand{\np}{Nucl.Phys. \underline}

\newcommand{\mev}{{\rm MeV}}

\begin{document}
\setcounter{page}{1}
\begin{flushright}
ROME prep. 94/981 \\  
18 February 1994.
\end{flushright}

\centerline{\LARGE{\bf{A High Statistics Lattice
 Calculation of $f^{static}_B$ }}}
\centerline{\LARGE{\bf{ at $\beta=6.2$ using
the Clover Action}}}
\vskip 0.5cm
\centerline{\bf{C.R. ALLTON$^1$, M. CRISAFULLI$^1$, V. LUBICZ$^1$,
G. SALINA$^{2}$, G. MARTINELLI$^{1}$,}}
\centerline{\bf{A. VLADIKAS$^{2}$}}
\vskip 0.2cm
\centerline{\bf{A. BARTOLONI$^1$, C. BATTISTA$^{1}$, S. CABASINO$^{1}$,
 N. CABIBBO$^{2}$,}}
\centerline{\bf{F. MARZANO$^{1}$, P.S. PAOLUCCI$^{1}$, J. PECH$^{1}$,
 F. RAPUANO$^{1,3}$,}}
\centerline{\bf{R. SARNO$^{1}$, G.M. TODESCO$^{1}$,
 M. TORELLI$^{1}$, W. TROSS$^{1}$,}}
\centerline{\bf{P. VICINI$^{1}$}}
\centerline{The APE Collaboration}

\vskip 0.5cm
\centerline{$^1$ Dip. di Fisica, Univ. di Roma \lq La Sapienza\rq and
INFN, Sezione di Roma,}
\centerline{P.le A. Moro, I-00185 Rome, Italy.}
\centerline{$^2$ Dip. di Fisica, Univ. di Roma \lq Tor Vergata\rq
and INFN, Sezione di Roma II,}
\centerline{Via della Ricerca Scientifica 1, I-00173 Rome, Italy.}
\centerline{$^3$ Theory Division, CERN, 1211 Geneva 23, Switzerland.}
\date{}
\begin{abstract}
{We present a calculation of $f_B$ in the
static limit, obtained by numerical simulation of quenched
QCD, at $\beta=6.2$ on a $18^3 \times 64$ lattice,
using the SW-Clover quark action.
The decay constant has been extracted by studying heavy(static)-light
correlation functions of different smeared operators,
on a sample of 220 gauge field configurations.
We have obtained $f_B^{static}=(290 \pm 15 \pm 45)$ MeV,
where the first error comes from the uncertainty in the determination
of the matrix element
and the second comes from the uncertainty in the lattice spacing.
We also obtain
$M_{B_s}-M_{B_d}=(70 \pm 10)$ MeV and
$f^{stat}_{B_s}/f^{stat}_{B_d}=1.11(3)$.
A comparison of  our results with other calculations of the
same quantity is made.}
 \end{abstract}
In the last few years, several groups\cite{almms}-\cite{fnalnew}
 have computed on the lattice the pseudoscalar
decay constant in the static limit, $f_B^{stat}$,
using the method originally proposed by E. Eichten\cite{eichtenseillac}.
One can combine  the information obtained from the calculation
of the decay constant in  a range of masses near the charm mass with
$f_B^{stat}$. In this way
 it is possible to reduce the uncertainty in the prediction
of $f_B$, which, on current lattices,
 can only be obtained by interpolation  in the heavy quark
mass\cite{labrenz,ukqcd,orsw}.  To improve the accuracy on
$f_B$ it is then important to reduce the uncertainties  in the static case.
These uncertainties are still rather large, as it can be seen from
the large spread of values reported in the
literature\cite{almms}-\cite{fnalnew}. They are mostly due to
lattice artefacts,  to the
calibration of the lattice spacing and to the renormalization of
the lattice axial current. \par To reduce lattice artefacts,
two groups\cite{ukqcd,apefb1}
 have found it convenient to use the SW-Clover improved action\cite{sheik}.
In a previous study we have calculated $f_B^{stat}$ at $\beta=6.0$,
using cubic smeared sources\cite{apefb1}. In this letter we extend the study to
a smaller lattice spacing, $a$, corresponding to $\beta=6.2$ (again using
the SW-Clover action).
In the future, a further reduction of discretization errors
will eventually be obtained by comparing  results at different values of $a$
and extrapolating to the continuum. This demands a very high
accuracy in the determination of $f_B^{stat}$
at any fixed value of the lattice spacing.
Our work is a step in this direction.
\par In ref.\cite{apefb1}, we found that,
by using  a cubic smearing source in the Coulomb gauge,
there is an optimal smearing size for which it is easiest to isolate
the lightest pseudoscalar state,
corresponding to the B-meson in the static limit. We confirm this result
and find that, using a double smeared source\cite{almms,apefb1}
at $\beta=6.2$, the optimal size is $L_s=11$
($L_s=13$ is the
optimal size for single smeared sources).
$L_s=11$ and double smeared sources have been used to obtain
the  results reported in the abstract and below\footnote{At
 $\beta=6.0$ the optimal smearing size was found to be $L_s=5-7$.
For a more detailed discussion see ref.\cite{apefb1}.}.
We will also compare different smearing sizes ($7 \le L_s \le 13$).
\par The decay constant in the static limit can be derived from
the matrix element of the static-light quark lattice axial current
$Z^L$ \footnote{The precise definition of $Z^L$ will be given in
eq.(\ref{eq:revi}).}:
\be
f_{P}^{stat} \sqrt{M_P} = \sqrt{2} \,\,
Z_A^{stat}(M_P a, \alpha_s) \,\, Z^{L} \, a^{-3/2} \label{eq:prima}
\ee
where $f_P^{stat}$ is the decay constant for a meson of mass $M_P$.
The renormalization constant $Z_A^{stat}(M_P a ,\alpha_s)$
includes the anomalous dimension
of the axial current in the effective theory and the $O(\alpha_s)$
correction necessary to relate the lattice operator in the
effective theory to the continuum operator in the full theory.
The estimate of $Z_A^{stat}$
is subject to a large uncertainty\cite{eh}-\cite{prossi}.
To evaluate it,
we have used the ``boosted'', tadpole improved perturbation theory proposed
in ref.\cite{lm1,lm2}. The uncertainty in the determination of $Z_A^{stat}$
is given by the uncertainty on the value of the
boosted coupling $g^2_V$ to be used and by higher order corrections.
We have used two different definitions of $g^2_V$, i.e. $g^2_V= (8
K_c)^4 6/\beta$
and $g^2_V= 6/\beta /<1/3 TrU_p>$\cite{labrenz,lm1,lm2}. With the SW-Clover
action, unlike the Wilson case, the two determinations of
$g^2_V$ almost coincide and we
get $Z_A^{stat}=0.805-0.812$\cite{pitbor,bhill}.
Thus we have chosen $Z_A^{stat}=0.81$.
Using $a^{-1}=(3.0\pm 0.3)$ GeV, as suggested by the calibration of the
scale from $M_\rho$ and $f_\pi$ from this
work and ref.\cite{ukqcd1}, together with $Z^L=0.111(6)$, we get
$f_{P}^{stat} \sqrt{M_P} = (0.66 \pm 0.04 \pm 0.10)\, {\rm GeV}^{3/2}$
corresponding to:
\be
f_B^{stat}=(290 \pm 15 \pm 45) \mev,
\ee
where
the first error  comes from the uncertainty in $Z^L$ and the second comes from
the uncertainty in the scale. When
the scale is taken from the string tension, $a^{-1} \sim 2.7$ GeV,
one gets a lower value, i.e. $f_B^{stat}=(245 \pm 15) $ MeV.
The calibration of the scale from the string tension is more convenient
when studying the scaling properties of $f_B^{stat}$ with $a$,
since the error on this quantity is of $O(a^2)$.
It is however difficult to express the lattice string tension in terms of
a well defined experimental quantity, in order
to obtain the absolute scale.
If we use the naive perturbative value of the renormalization
constant, ie. $Z_A^{stat}=0.88$, and use $a^{-1} = (3.0 \pm 0.3)$ GeV,
we obtain instead $f_B^{stat}=(315 \pm 15 \pm 45)$ MeV.
\par Most of the methods and techniques used in the present study have been
developed and explained in the literature on the same
subject\cite{almms}-\cite{fnalnew}.
Details can be found for example in a previous publication
by our collaboration\cite{apefb1}.
In the following we explain only those aspects
which are new with respect to ref.\cite{apefb1}.
The main differences are the following:
\par 1) smearing in the origin;
\par 2)  calculation of the matrix elements of the local axial current,
$Z_L$, by fitting the smeared-local correlation function;
\par 3) a method to extract $Z_L$ from ratios of correlation functions
and without any fitting procedure.
\par
In this study, we have used local-smeared correlation functions
with the smeared source in the
origin. This strongly reduces the statistical noise in the
correlators\cite{eichtencapri}.
In our previous work at $\beta=6.0$ we used instead the smeared-local
correlations with the smeared source in the sink. To improve
the results of ref.\cite{apefb1}, we are planning
to repeat the calculation at $\beta=6.0$  with the smeared source in
the origin. Points 2) and 3) will be discussed below.
 \par
We have generated 220 configurations in the quenched approximation, at
$\beta=6.2$ on a $18^3 \times 64$ lattice. The configurations
were obtained from several independent runs, each with
 a  thermalization of 3000 Montecarlo sweeps (5 hits),
starting from a cold configuration. Independent configurations
were  gathered every 800 sweeps.
We have then transformed each configuration to the Coulomb
gauge, with the same accuracy as in ref.\cite{apefb1}.
We have used the $O(a)$ improved SW-Clover action\cite{sheik}.
The  operators appearing in the correlation functions
  are  ``improved"
by modifying the light quark propagators as explained in sec.9 of
 ref.\cite{mssv}.
The statistical errors quoted in this letter have been estimated using
the jacknife method, by decimating 10 configurations at a time.
In ref.\cite{apefb1} we verified that this gives a reliable estimate of the
 errors.
Throughout this letter we have computed light quark correlation functions
using a ``thinning" trick. This method consists of the
following: one only  considers  one point out
of three in each direction when summing the correlation
functions over space, at fixed time. This allows a saving of
 a factor of 27 in the
memory required for the quark propagators. Thinning was a necessity imposed
by memory limitations of the APE-tubo\cite{fred91} used for the present
simulation. We have checked that the effect of thinning on 2-point
correlation functions is totally negligible\cite{thin}.
\par We have inverted the light quark propagators for three
 values of the bare quark
masses, $K=0.14144, 0.14190, 0.14244$.
For comparison, the first $K$ value has
been chosen equal to one of those used by the UKQCD
collaboration\cite{ukqcd},
at the same value of $\beta$.
 Our lightest quark mass is slightly larger than in
ref.\cite{ukqcd} since, for lack of memory, we have to work on a
smaller space volume ( the UKQCD Collaboration is working
at $\beta=6.2$ on a $24^3 \times 48$ lattice).  \par
We have  measured the pseudoscalar ($M_5$) and vector ($M_V$) meson masses
and the matrix elements
$\sqrt{Z_5} = < 0 \vert \bar \psi \gamma_5 \psi \vert P>$ and
$\sqrt{Z_V} = < 0 \vert \bar \psi \gamma_k \psi \vert V>$
by fitting  the correlation functions of the local
pseudoscalar density and local vector current with components $k=1,2,3$,
in the time interval $15 \le t \le 28$. By studying several time
intervals, we have checked that the contamination due to higher mass
excitations
induces a negligible systematic error  on the masses
and  the matrix elements of the local operators,
in the range of quark masses explored in this study.  The results
at each $K$ and $K_c$ ( the critical value of the hopping
parameter) are reported in table \ref{tab:masses}.
 From a linear fit in 1/K to $M_5^2$ we obtain $K_c=0.14315(4)$.
Using $M_K=498$ MeV and the scale from the rho mass we obtain
$K_s=0.1422(2)$, where $K_s$ is the Wilson parameter corresponding
to the strange quark.
\begin{table}
\begin{center}
\begin{tabular}{||c|c|c|c|c||} \hline \hline
K                   & 0.14144   & 0.14190   & 0.14244   & $K_c$     \\ \hline
\hline
$M_5$               & 0.292(3)  & 0.247(3)  & 0.187(4)  & $-$       \\ \hline
$M_V$               & 0.375(6)  & 0.338(8)  & 0.291(18) & 0.238(21) \\ \hline
$Z_5\times 10^3$    & 8.55(44)  & 7.43(46)  & 6.65(60)  & $-$       \\ \hline
$Z_V\times 10^3$    & 1.80(18)  & 1.28(18)  & 0.76(27)  & $-$       \\ \hline
$f_P/Z_A^{Ren}$     & 0.063(2)       &0.059(2)       &0.052(3)       &
0.045(4)       \\ \hline
$f_P/Z_A^{Ren} M_V$ & 0.169(5)  & 0.173(7)  & 0.180(15) & 0.186(20) \\ \hline
$1/Z_V^{Ren} f_V$   & 0.302(7)  & 0.313(10) & 0.326(22) & 0.341(27) \\ \hline
\hline
\end{tabular}
\caption{ \it{Pseudoscalar and vector meson masses
and matrix elements  at different values
of $K$. The corresponding values extrapolated linearly
to $K_c$  are also given.}}
\label{tab:masses}
\end{center}
\end{table}

The physical size of the
lattice, obtained from the vector meson mass extrapolated linearly
in $M_5^2$ to the chiral limit, is
$a^{-1} = (3.2 \pm 0.3) $ GeV.
The pseudoscalar ($f_P$) and vector ($1/f_V$)
decay constants have  been
extracted using standard methods,  see for example ref.\cite{apefb1},
from the local vector and axial vector currents. In table \ref{tab:masses}
we report the lattice values of $f_P/ Z_A^{Ren}$,
 the ratio ${f_P/ Z_A^{Ren} M_V }$
and $1/Z_V^{Ren}f_V$  as a function of the light quark mass.
$Z_{V,A}^{Ren}$ are the renormalization constants of the
vector and axial vector lattice currents\cite{smit}-\cite{boc} which,
for the SW-Clover action, have been computed non-perturbatively at
$\beta=6.2$,
 $Z_V^{Ren}=0.831(2)$ and $Z_A^{Ren}=1.04(1)$\cite{marta}.
We have  extrapolated ${f_P/ Z_A^{Ren}}$, ${f_P/ Z_A^{Ren}M_V }$
and $1/ Z_V^{Ren}f_V$ linearly
in $M_5^2$ to the chiral limit, obtaining the numbers in
table \ref{tab:masses}.
Using the non-perturbative determinations of $Z_V^{Ren}$ and $Z_A^{Ren}$
given above, and  $M_{\rho} = 770 \, {\rm MeV} $ we get
\be f_\pi = (149 \pm 16)\, \mev ~,~~~~~~~~~~
     \frac{1}{f_\rho} = 0.28 \pm 0.02 \ee
We also obtain
$ f_{K} / f_{\pi} -1 = 0.11 \pm  0.04$ and
   $ 1/f_\phi =  0.27 \pm 0.02$.
We give for comparison the experimental values: $f_\pi=132$ MeV,
$1 /f_{\rho}  =  0.28$,  $f_K / f_{\pi}-1= 0.22$
and $1/f_\phi  =  0.23$. The slope of the pseudoscalar decay constant
as a function of the quark mass is lower than what we found in previous
calculations at $\beta=6.0$ and for this reason we get a rather low value
for $f_K/f_\pi-1$. We believe however
that this effect will disappear with better
statistics. We can also use $f_\pi$ to fix the scale. By using the value
of $Z_A^{Ren}$ quoted above we obtain $a^{-1}=(2.8 \pm 0.3)$ GeV to be compared
to the value obtained from $M_\rho$ of $a^{-1}=(3.2 \pm
0.3)$ GeV\footnote{From
the axial current-axial current two point function one can also get
 a lattice estimate of $f_\pi$.
In this case, using the same $Z_A^{Ren}$ and the experimental value of $f_\pi$
we also get $a^{-1} = (2.8 \pm 0.3)$ GeV.}.
At $\beta=6.0$, we found that the scale determined from
$f_\pi$ and $M_\rho$ were in very good agreement. In the
present case there is a difference of about $400$ MeV that we cannot
resolve, given the large statistical errors.
We also obtain $M_K^* = (893 \pm 7)$ MeV and $M_\phi = (1020 \pm 13)$ MeV
which compare very well with the experimental values of $892$ MeV
and $1019$ MeV respectively.
\vskip 0.3 cm
The static B meson decay constant, $f^{stat}_{B}$
has been obtained from the correlation functions of
three different currents:
\be
A^{L}_{\mu}(x) = \bar Q(x) \gamma_{\mu}\gamma_{5} q(x) \label{lc}
\ee
\be A^{S}_{\mu}(x) = \sum_i \bar Q(x_i) \gamma_{\mu}\gamma_{5} q(x)
\label{sc} \ee
\be A^{D}_{\mu}(x) = \sum_{i,j} \bar Q(x_i) \gamma_{\mu}\gamma_{5} q(x_j)
 \label{dc} \ee
\noindent
where the sum over $i$ ($j$) is extended over a  spatial cube of size $L_s$,
centered at $x$ and the heavy quark $Q$ is taken from the static limit.
We have computed the correlation
functions of the above currents, $C^{LL}$, $C^{LS}$,
$C^{LD}$, $C^{SS}$
and $C^{DD}$,
  for  $L_s=7,9,11,13$. $C^{LD}$ is defined as:

\begin{equation}
        C^{LD}(t) = \sum_{\vec x}<0|A^{L}_{0}(\vec x,t)
A^{D}_{0}(\vec 0,0)^{\dag}|0>
\end{equation}
and similarly for the others.
At large time distances the correlation functions behave as the propagator
of a single state:
\begin{equation} C^{LD}(t) = Z^L Z^D e^{- \Delta E t}
 \label{eq:revi} \end{equation}
The constants $Z^L,Z^S$ and $Z^D$ are related to the  matrix
elements of the different axial currents
 and $\Delta E$  is the binding energy of the lightest pseudoscalar
state.

\par  A standard method  to extract $Z^L$ is the
following:
\par
1) At large time distances the correlation
functions are dominated by the lightest propagating state.
Let us call   $t_i$ the time at which we start observing
a plateau both for
the effective binding energy:
\be \Delta E_{DD}^{eff} = log [ C^{DD}(t) / C^{DD}(t+1)] \label{one}\ee
and for the ratio:
\begin{equation} R(t) = {C^{LD}(t) \over C^{DD}(t)} \rightarrow \frac
{Z^L}{Z^D}
\label{two} \end{equation}
 $t_i$ clearly depends on the smearing size and it is determined
by observing that $\Delta E_{DD}$ and $R$ do not vary appreciably
from time to time.
\par 2) For $t \ge t_i$, i.e. in the time
interval where (\ref{one}) and (\ref{two}) have
a plateau, we fit the two-point smeared-smeared correlation
$C^{DD}$ to the expression:
\begin{equation} C^{DD}(t) = (Z^D)^2 e^{- \Delta E_{DD} t}
\label{eq:uno} \end{equation}
{}From this fit we obtain the matrix element of the corresponding axial
current,
 $Z^D$.
\par 3) We then compute $R$  defined as:
\be R=  \frac {\sum_{t=t_i}^{t_f} R(t)/ \sigma_{ R(t)}^2}
{\sum_{t=t_i}^{t_f} 1/\sigma_{ R(t)}^2} \label{rt} \ee
 where
$\sigma_{ R(t)}$ is the jacknife error on the ratio $R(t)$ at the time $t$.
We used $t_f=15$ for which the errors in the static-light correlation
functions are reasonable. 
\par
4) $Z^L$ is then obtained from the product:
\be Z^L = R \times Z^{D} \ee
where $Z^D$ is found from the fit  of the correlation
function to the expression in  eq.(\ref{eq:uno}).
We will call this method {\bf DD-mass fit}.
\par We now give two other methods that we have used in this letter
to determine $Z^L$.
\par i) {\bf LD-mass fit}: in point 2) we replace  the $C^{DD}$ fit with a fit
of $C^{LD}$. In this way we obtain
the binding energy and $P=Z^L Z^D$, i.e. the products of the matrix elements
of the local and smeared currents. We then compute $Z^L$ from the product
of $P$ with $R$ as defined in 3), $Z^L = \sqrt{R \times P}$.
The advantage of this method is that $C^{LD}$ has the smaller statistical
errors. A possible disadvantage is that the effective binding energy
for $C^{LD}$ approaches
its asymptotic value at large times from below, contrary to the case in which
one uses $C^{DD}$. However the quality of the plateau that we observe
in the effective mass $\Delta E^{LD}$, from $C^{LD}$,
 is such that we do not doubt that we have isolated  the lightest state.
This is shown in fig.\ref{fig:emass} where $\Delta E^{LD}$ and
$\Delta E^{DD}$  are shown together  (see  also table
\ref{tab:emass}). For comparison on the quality
of our data we also report the results of the UKQCD collaboration,
at the same value of $\beta$ and $K$, obtained with a statistics of 20
configurations on a $24^3 \times 48$ lattice. We will compare our
static results with those of ref.\cite{ukqcd}
and our light-light results with ref.\cite{ukqcd1} at the end of this letter.
\par ii) {\bf Ratio method}:
 we compute the following ratio of correlation functions:
\be R_{Z^L}(t_1,t_2)=\frac{C^{LD}(t_1)C^{LD}(t_2)}{C^{DD}(t_1+t_2)}
\ee
At large $t_1$ and $t_2$ ($t_{1,2} \ge t_i$), it is clear from
eqs.(\ref{eq:revi},\ref{eq:uno}) that $R_{Z^L}\rightarrow (Z^{L})^2$.
In practice, we average
$R_{Z^L}(t_1,t_2)$ from $t_1+t_2=t_m=2 t_i$ to
a certain  maximum value $t_1+t_2=t_M$, for which the errors are
reasonably small,
with  all $t_{1,2}$ which satisfy  $ t_{1,2} \ge t_i$.
The advantage of this method is that $Z^L$ is obtained directly
and no fitting is necessary. The possible disadvantage is that we have
to compute $C^{DD}$ at large time distance, i.e. $\ge 2 t_i$.
We note here that this method can be applied
not only to measurements of $f_P^{stat}$, but to a large variety of matrix
elements.
\begin{figure}[t]   
    \begin{center}
       \setlength{\unitlength}{1truecm}
       \begin{picture}(6.0,9.0)
       \end{picture}
    \end{center}
    \caption[]{\it{$\Delta E_{DD}^{eff}$
and $\Delta E_{LD}^{eff}+0.2$ from $C^{DD}$ and $C^{LD}$ as a function
of the time $t$ for $L_s=11$ and $K=0.14144$. We also report the published data
of the UKQCD collaboration\cite{ukqcd}. The points of ref.\cite{ukqcd}
have been moved by $-0.2$ in order to distinguish them from
our results. The lines joining the
points are there to help the reader distinguish the different cases.}}
    \protect\label{fig:emass}
\end{figure}
%
\begin{figure}[t]   
    \begin{center}
       \setlength{\unitlength}{1truecm}
       \begin{picture}(6.0,9.0)
       \end{picture}
    \end{center}
    \caption[]{\it{$\Delta E_{DD}^{eff}$
and $\Delta E_{LD}^{eff}+0.2$ from $C^{DD}$ and $C^{LD}$ as a function
of the time $t$ for $L_s=7-13$ and $K=0.14144$. We also give best estimate,
$\Delta E^{eff}=0.561(3)$, the band of which appears as a thick line
in the figures. Notice that  $L_s=11$ works well both for $C^{LD}$ and
$C^{DD}$ at moderate values of $t$. }}
    \protect\label{fig:deplateau}
\end{figure}
%
\par Several consistency checks are possible.
 One should find  a plateau corresponding to the same
value of the effective mass for the
$C^{LD}$ and $C^{DD}$. Since we extract $Z^L$ by taking ratios,
we also need to overlap the
time ranges in which the plateau in the effective mass is observed.
We show that this is possible in
figs.\ref{fig:emass},\ref{fig:deplateau} where we give
$\Delta E_{DD}^{eff}$ and
$\Delta E_{LD}^{eff}$ as a function of the time for $L_s=7-13$.
\par
A more quantitative check of the reliability of the results is given
by the fact that different correlation functions give the same
effective binding energy.
The results and errors for $\Delta E_{DD}^{eff}$ and
$\Delta E_{LD}^{eff}$ with different smearing sizes and time intervals,
which we consider in the good range of $L_s$ and $t$, are reported in table
\ref{tab:emass}.
The binding energies
extrapolated linearly in $M_5^2$
to the chiral limit are also given in the same table.  From the dependence
of the binding energy  on the quark mass one can derive the mass difference
$M_{B_s}-M_{B_d}$\cite{boc2}.
Our best estimate for this quantity is
$M_{B_s}a-M_{B_d}a = 0.020(4) $, ie. $M_{B_s}-M_{B_d} = 70(10)$ MeV.
We can see from fig.\ref{fig:deplateau} and table \ref{tab:emass}
that the smearing with the best plateau
both in the $DD$ and $LD$ case is $L_s=11$.
 Consequently all the results
given for $f_P^{static}$ have been obtained with this
smearing size and $t_i=5$ and $7$.
\par
A further quantitative check is that the three different methods give
the same value of $Z^L$. A comparison of the different determinations
can be done from table \ref{tab:zl}. $Z^L$,
extrapolated linearly in $M_5^2$
to the chiral limit, is also given in the same table.  From
the dependence of $Z^L$ on the quark mass we have obtained
the value of $f_{B_s}/f_{B_d} = 1.11(3)$ (see \cite{apefb1}).
Finally, we have checked that the $S-$smearing (see eq.(\ref{sc})),
when we use $L_s=13$,
gives consistent results to the $D-$smearing (eq.(\ref{dc})),
thus giving confidence in our methods.
\par
{}From the results for the effective mass and $Z^L$ at different values
of the Wilson parameter and in the chiral limit, other groups working
on the same problem and using the same quark action, but
different smearing techniques to isolate the lightest state, can check our
results.
Our best estimate of $Z^L$ is $0.111(6)$ which we obtain from a
weighted average of the results in
table \ref{tab:zl}, where the error includes the spread of values in the table
added in quadrature with the statistical error.
\par
Once $Z^L$ is established for a given value of the $\beta$,
the determination of $f_B^{stat}$ still depends upon several factors.
Apart from effects of $O(a)$, which we expect to be smaller with the
SW-Clover action, there are the uncertainties coming from the calibration of
the lattice
spacing, $a$, and from the renormalization constant of the axial current
in the static theory (see eq.(\ref{eq:prima})).
\begin{table}
\begin{center}
\begin{tabular}{|c|c|c|c|c|} \hline
\multicolumn{1}{|c}{} & \multicolumn{4}{|c|}{$\Delta E_{DD}$
}
 \\ \hline
K &  $L_s=11$ $t=5$-$15$ &  $L_s=11$ $t=7$-$15$ &
 $L_s=13$ $t=5$-$15$&  $L_s=13$ $t=7$-$15$ \\
\hline
0.14144 & 0.560(5) & 0.556(7)&0.563(6)&0.560(8)\\
0.14190 & 0.548(6)& 0.545(7)&0.551(7)&0.548(8)\\
0.14244 & 0.532(6)& 0.529(8)&0.535(8)&0.531(10)\\
\hline
$K_c$  & 0.514(6) &0.511(9) &0.515(7) &0.513(10)\\
\hline
\hline
\multicolumn{1}{|c}{} & \multicolumn{4}{|c|}{$\Delta E_{LD}$
}
 \\ \hline
K &  $L_s=11$ $t=5$-$15$ &  $L_s=11$ $t=7$-$15$ &
 $L_s=13$ $t=5$-$15$ &  $L_s=13$ $t=7$-$15$ \\
\hline
0.14144 & 0.562(2) & 0.561(3)&0.550(2)&0.553(4)\\
0.14190 & 0.552(3)& 0.551(3)&0.540(3)&0.543(4)\\
0.14244 & 0.541(3)& 0.539(3)&0.528(3)&0.531(5)\\
\hline
$K_c$  & 0.526(4) &0.524(5) &$-$&$-$\\
\hline
\end{tabular}
\caption{ \it{$\Delta E_{DD}$ and $\Delta E_{LD}$
 for $L_s= 11$ and  $13$. The fits  done in different
time intervals have been reported in order to show the stability of the
results. The agreement is very good with the only exception of
$\Delta E_{LD}$ when $L_s=13$. The reason can be understood by looking at
fig.\ref{fig:deplateau}
where it is shown that there is a drift
in the effective mass as a function of the time. Analogous problems
are encountered with $C^{LD}$ for $L_s=7$ and $9$. On the contrary
this table shows that with $L_s=11$ one gets results consistent,
within very tiny errors, from both $C^{LD}$ and $C^{DD}$ and different
time intervals.}}
\label{tab:emass}
\end{center}
\end{table}

\begin{table}
\begin{center}
\begin{tabular}{|c|c|c|} \hline
\multicolumn{1}{|c}{} &
\multicolumn{2}{|c|}{$Z^L$ from the {\bf DD-mass fit}}
 \\ \hline
K &  $L_s=11$ $t=5$-$15$ &  $L_s=11$ $t=7$-$15$ \\
\hline
0.14144 & 0.134(3) & 0.132(4) \\
0.14190 & 0.127(3)& 0.125(4)\\
0.14244 & 0.119(3)& 0.116(4)\\
\hline
$K_c$   & 0.108(4)& 0.104(5)\\
\hline
\hline
\multicolumn{1}{|c}{} &
\multicolumn{2}{|c|}{$Z^L$ from the {\bf LD-mass fit}}
 \\ \hline
K &  $L_s=11$ $t=5$-$15$ &  $L_s=11$ $t=7$-$15$ \\
\hline
0.14144 & 0.136(2) & 0.135(2) \\
0.14190 & 0.130(2)& 0.129(2)\\
0.14244 & 0.123(2)& 0.121(2)\\
\hline
$K_c$   & 0.114(2)& 0.111(3)\\
\hline
\hline
\multicolumn{1}{|c}{} &
\multicolumn{2}{|c|}{$Z^L$ from the {\bf Ratio method}}
 \\ \hline
K &  $L_s=11$ $t=10$-$15$ &  $L_s=11$ $t=14$-$15$ \\
\hline
0.14144 & 0.135(2) & 0.134(5) \\
0.14190 & 0.129(2) & 0.128(5) \\
0.14244 & 0.121(2) & 0.119(6) \\
\hline
$K_c$   & 0.111(2)   & 0.109(7) \\
\hline
\hline
\end{tabular}
\caption{ \it{$Z^L$ obtained from the three methods (see text)
using $L_s= 11$.
The fits  are done in  two time intervals ($t=5-15$ and $t=7-15$)
in order to show the stability of the  results.
In the {\bf ratio method} we have chosen $t_m=10$ and $t_M=15$ for
$t_i=5$ or
$t_m=14$ and $t_M=15$ for $t_i=7$, see text.
The agreement between all three methods is quite satisfactory.}}
\label{tab:zl}
\end{center}
\end{table}
\par There is only one group which has worked at this value of $\beta$,
using the SW-Clover action, UKQCD\cite{ukqcd,ukqcd1}. We now compare our
results
with theirs  for the light meson sector and the static case.
\begin{figure}[t]   
    \begin{center}
       \setlength{\unitlength}{1truecm}
       \begin{picture}(6.0,9.0)
       \end{picture}
    \end{center}
    \caption[]{\it{$M^2_5$, $M_V$, $f_P/Z_A^{Ren}$ and
$f_P/(Z_A^{Ren}M_V)$ as a function of $1/K$. We plot both
our  and the UKQCD\cite{ukqcd1} results, the latter obtained
with a statistics of 60 configurations. Only the statistical errors
are shown. The lines are linear fits in $1/K$ to our data.}}
    \protect\label{fig:comparison}
\end{figure}

In figs.\ref{fig:comparison} we plot $M_5^2$ and $M_V$ as a function
of $1/K$.  Our results are in good agreement for $M_5^2$
(we find essentially the same $K_c$) but are systematically
lower for $M_V$. In fact, at the only value of $K$ which is common to the two
calculations, our result coincides with the value  of $M_V$
reported as $M_V^{low}$ in table 1
of ref.\cite{ukqcd1}. The results are in better agreement if
we include in the comparison
 the systematic error quoted in ref.\cite{ukqcd1}, which was  estimated
by looking at the different determinations of $M_V$, obtained by varying
the time interval of the fits.  A possible
origin of the difference is that UKQCD fits the
correlation functions from $t_i=13$ to $23$.
The difference in $M_V$ is reflected in
the estimates of $a^{-1}$, the inverse lattice spacing,
which in their case is $a^{-1} = (2.7 \pm 0.1)$ GeV
(no systematic error is reported) to be compared with our value of
$a^{-1} = (3.2\pm 0.3)$ GeV.
On the contrary, our result for $f_P$ is systematically above
the result of ref.\cite{ukqcd1}, see also table \ref{tab:masses}
and figs. \ref{fig:comparison}.
However the results are in this case compatible
within the statistical fluctuations.
The combined effect of a smaller $M_V$ and a larger $f_P$ is that
the difference between the two groups
  becomes a discrepancy for $f_P/M_V$\footnote{Note that
we find that $f_P/M_V/Z^{Ren}_A$ is slightly increasing in the chiral
limit, contrary to previous findings at $\beta=6.0$\cite{apefb1}.
This is probably due to a statistical effect.}.
 From the above discussion we have seen that most of this is due to $M_V$.
\par
For the static case, at K=0.14144, our results
for the binding energy are definitely incompatible with those of
UKQCD\cite{ukqcd}. They quote $\Delta E^{eff}=0.59 \pm 0.1$ to be compared
with our results, given in table \ref{tab:emass}. At this point a comparison
of $Z^L$ becomes difficult. In any case, the difference for $Z^L$ is not too
large: they find $Z^L=0.142^{+7}_{-6}$ at K=0.14144, to be compared with
a typical value (error) for our results of $Z^L=0.136(2)$,
and $Z^L=0.124^{+8}_{-7}$ at $K_c$, to be compared to our estimate
of $Z^L=0.111(6)$.
\par
A prerequisite to a precise determination of $f_B^{stat}$ demands
agreement on the value  of $\Delta E^{eff}$
 and  $Z^L$ obtained with a given action at a certain value
of $\beta$. Our group and the UKQCD collaboration have decide to work with the
SW-Clover action in order to reduce lattice artefacts.  In this paper we have
presented a very accurate determination of $Z^L$, derived with three different
methods, which give compatible results, within tiny statistical errors.
Our results for the binding energy  show some difference with
the results of ref.\cite{ukqcd}
which, by taking the quoted statistical errors, appear to be of systematic
origin. The same can be said for the mass of the vector meson \cite{ukqcd1}.
Other differences, for example $f_P$, can instead be the effect of
statistical fluctuations.

\section*{Acknowledgments}
\par
We acknowledge C. Sachrajda for interesting discussions
and the members of the UKQCD collaboration who have provided us with
their results for comparison. We acknowledge
the partial support of the MURST, Italy, and  the INFN.

\end{document}